\documentclass[
reprint,
twocolumn,
]{revtex4-1}

\usepackage{graphicx}
\usepackage{subfigure}
\usepackage{amsmath}
\usepackage{amssymb}
\usepackage{epsfig}
\usepackage{color}
\usepackage{amsfonts}
\usepackage{wrapfig}
\usepackage{epstopdf}

\begin{document}
\title{Cu codoping control over magnetic precipitate formation in ZnCoO nanowires}

\author{S.~Granville} \email{simon.granville@vuw.ac.nz}
  \affiliation{Institute of Condensed Matter Physics, Ecole Polytechnique F\'{e}d\'{e}rale de Lausanne-EPFL, 1015 Lausanne, Switzerland}
	\altaffiliation{now at Robinson Research Institute, Victoria University of Wellington, PO Box 33436, Lower Hutt 5046, New Zealand}
  \affiliation{The MacDiarmid Institute for Advanced Materials and Nanotechnology, Wellington, New Zealand}
\author{E.~Matei} 
  \affiliation{National Institute of Materials Physics, Atomistilor Street 105 bis, Magurele, Ilfov 77125, Romania}
\author{I.~Enculescu}
  \affiliation{National Institute of Materials Physics, Atomistilor Street 105 bis, Magurele, Ilfov 77125, Romania}
\author{Maria Eugenia Toimil-Molares}
 \affiliation{Gesellschaft f\"{u}r Schwerionenforschung (GSI), Planckstra{\ss}e 1, D-64291 Darmstadt, Germany}

\begin{abstract}
Using electrodeposition we have grown nanowires of ZnCoO with Cu codoping concentrations varying from 4-10 at.~$\%$, controlled only by the deposition potential. We demonstrate control over magnetic Co oxide nano-precipitate formation in the nanowires via the Cu concentration.  The different magnetic behavior of the Co oxide nano-precipitates indicates the potential of ZnCoO for magnetic sensor applications.  

\end{abstract}

\maketitle

The potential to combine magnetism with semiconducting properties in a single material is a major driving factor behind the recent intense interest into doping magnetic elements into well-known semiconductor materials, such as adding transition metal ion Co into wide bandgap semiconductor ZnO~\cite{Dietl2000}.  
More recently, attention has focused around the prospect of controlling the magnetic properties of such dilute magnetic semiconductors (DMS) through codoping with non-magnetic elements such as Ga~\cite{He2008}, Al~\cite{Liao2007} or Cu~\cite{Lin2004}.  Codoping can optimize the charge carriers for enhanced ferromagnetism~\cite{Spaldin2004,Lee2007} and has been touted as a way to modify the tendency of the magnetic dopant to cluster into magnetic nano-precipitates~\cite{Zhang2008}. While the wide focus in DMS has been on achieving robust ferromagnetism above room-temperature~\cite{Ueda2001,Schwartz2004,Kittilstved2005} \textit{without} such nano-precipitates~\cite{Norton2003,Deka2005,Mandal2006}, extremely large magnetoresistances~\cite{Tian2008} reported in such doped materials indicate a useful future as magnetic sensor elements. For such a purpose, the presence of magnetic nano-precipitates in a non-magnetic host matrix can be desirable~\cite{Jamet2006,Serrano-Guisan2006}.  Moreover, magnetic clusters may form unintentionally~\cite{boubekri_annealing_2009} at the temperatures used to process DMS and their presence may be missed unless advanced characterization techniques are used~\cite{lee_study_2004,ney_advanced_2010}.  Hence, it would be highly useful to understand how to actively control the level of nano-precipitate formation in DMS through codoping.

A theoretical study by Zhang \textit{et al.} predicted that Cu codoping would suppress the aggregation of Co in ZnO~\cite{Zhang2011}. By contrast, Pan \textit{et al.} recently showed that Co clusters form in thin films prepared under high temperature conditions with Cu codoping of above 2\%~\cite{Pan2012}.  In this letter, we demonstrate control over the formation of magnetic precipitates in electrodeposited ZnCoO nanowires through Cu codoping.  As a result, we are able to produce arrays of nanowires with differing magnetic properties using a simple and low-temperature growth process.

Out of the methods of preparation of nanowires, the template approach is one of the most widely used. It is relatively simple and allows for a very precise control of the morphology of the nanostructures.  Porous templates with channel diameter 200~nm were fabricated by GeV heavy ion irradiation of 30~$\mu$m thick polycarbonate foils (Makrofol N Bayer AG) and subsequent chemical etching~\cite{Enculescu2008}. Prior to etching, the foils were exposed to UV light from both sides to reduce the channel diameter distribution~\cite{cornelius_nanopores_2010}, resulting in a pore diameter variation of less than 10$\%$.  The electrochemical bath contains Zn(NO$_3$)$_2$ and different concentrations of Co(NO$_3$)$_2$, Cu(NO$_3$)$_2$ and CoCl$_2$ as the sources of the Co and Cu ions as well as polyvinylpyrrolidone (PVP) as an additive to improve pore wetting.  The bath was heated to 90$^{\circ}$C during deposition to ensure the nanowires were deposited with a good crystalline structure.

The preferred crystallographic orientation of the nanowire arrays was investigated by X-ray diffraction (XRD).  Magnetization measurements were made with a Quantum Design MPMS5 SQUID magnetometer, at temperatures from 2.5-350~K, and in applied fields of up to 50~kOe.  The magnetic measurements were made on nanowires still contained within the diamagnetic polymer membrane, with field applied both parallel and perpendicular to the nanowire length.  For further morphological and compositional characterization by scanning electron microscopy (SEM) and energy dispersive x-ray analysis (EDX), the polymer membranes were first dissolved in chloroform or dichloromethane and the nanowires were then transfered onto Si substrates.  

\begin{figure}
\includegraphics[width=8cm]{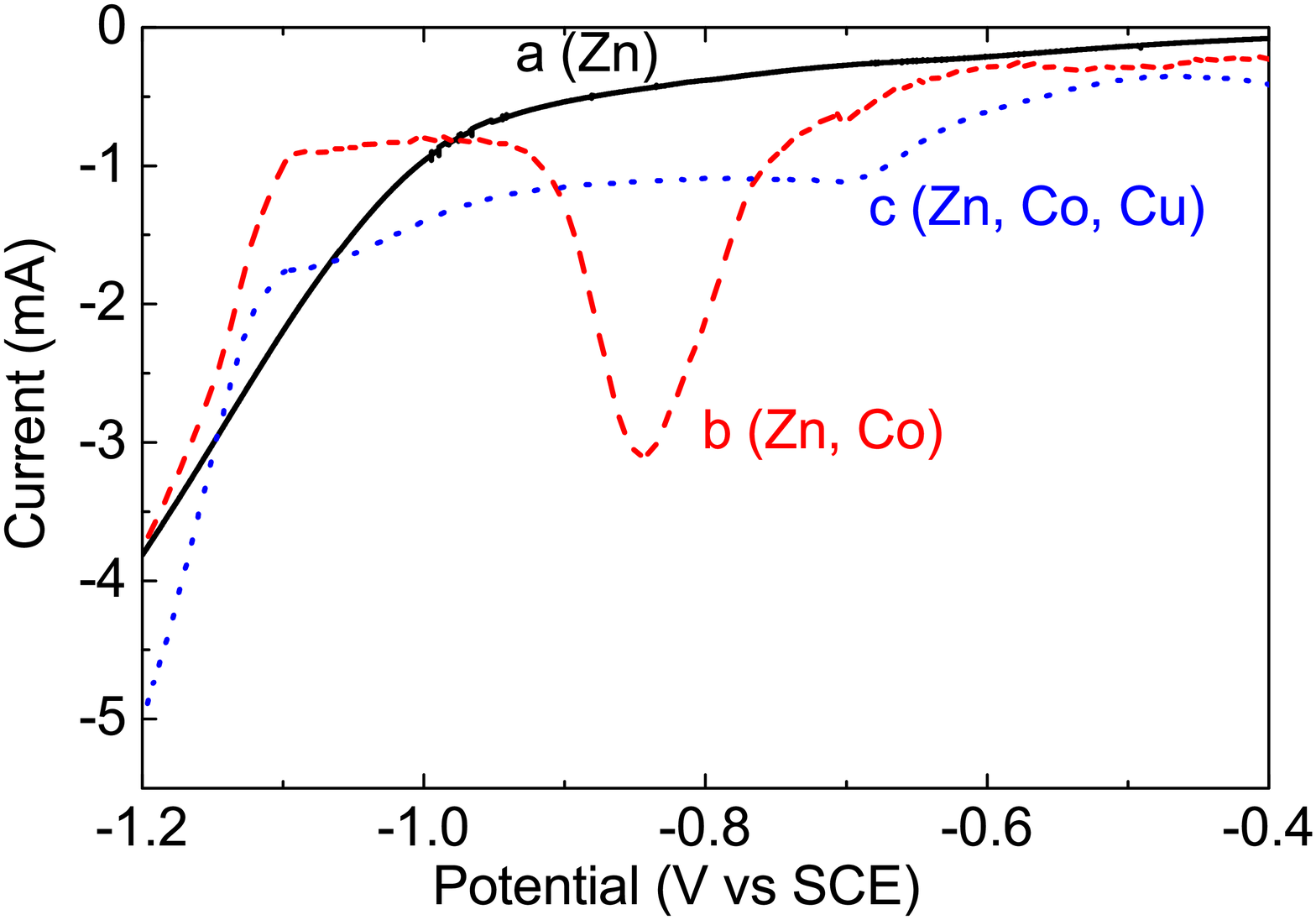}\vspace{-0.4 cm}
\caption{(color online) Electrochemical polarization curves for solutions containing (a) 0.05 M Zn(NO$_3$)$_2$, 1 g/l PVP ; (b) 0.05 M Zn(NO$_3$)$_2$ , 0.05 M CoCl$_2$, 1 g/l PVP ; (c) 0.05 M Zn(NO$_3$)$_2$, 0.05 M CoCl$_2$ and 0.1 mM Cu(NO$_3$)$_2$, 1 g/l PVP. 
\label{CV}}
\end{figure}

\begin{table*}[]\hspace{-6.0 cm}
\caption{\label{Compos}Electrochemical bath compositions, deposition potentials and resulting nanowire compositions.}
\begin{ruledtabular}
\begin{tabular}{l|cccccccccc}
Sample & Deposition & \multicolumn{5}{c}{Electrochemical bath composition} & \multicolumn{3}{c}{Zn$_{1-x-y}$Co$_x$Cu$_y$O nanowire}\\
number & potential & \multicolumn{5}{c}{Moles} & \multicolumn{3}{c}{composition} \\

 & (mV) & & Zn(NO$_3$)$_2$ & Co(NO$_3$)$_2$ & CoCl$_2$ & Cu(NO$_3$)$_2$ & & x (\%) & y (\%)  \\
\hline
920 (Co-only) & -750 & & 0.05 & 0.05 & - & - & & 4 & - \\
1136 (high Cu codoped) & -800 & & 0.05 & - & 0.05 & 1$\times$10$^{-4}$ & & 4.7 & 9.9 \\
1137 (low Cu codoped) & -900 & & 0.05 & - & 0.05 & 1$\times$10$^{-4}$ & & 4 & 4.6 \\
1138 (moderate Cu codoped) & -1000 & & 0.05 & - & 0.05 & 1$\times$10$^{-4}$ & & 3.6 & 8 \\
\end{tabular}
\end{ruledtabular}
\end{table*}

First, a solution containing 0.05 M Zn(NO$_3$)$_2$ and 1 g/L PVP was prepared, and a portion of this solution was then used to make separate solutions containing Co or both Co and Cu ions.  Figure~\ref{CV}(a) shows the electrochemical polarization curve recorded during reduction inside the membrane channels of a solution containing no Co and no Cu ions. Figures~\ref{CV}(b) and \ref{CV}(c) show the polarization curves measured on identical membranes using Co and (Co,Cu) containing electrolytes, respectively.

\begin{figure}
\includegraphics[width=8cm]{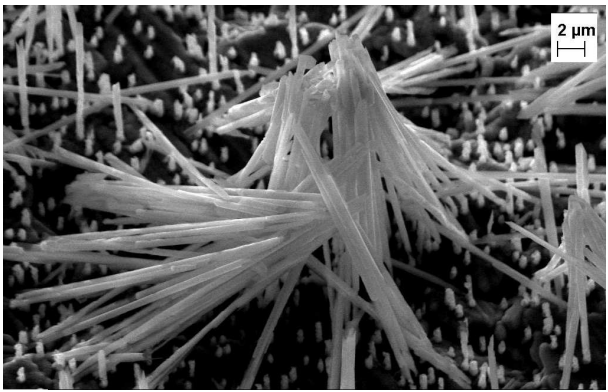}
\caption{SEM image of doped ZnO wires liberated from the template. The wires were deposited at -750 mV from a solution of: 0.05 M Zn(NO$_3$)$_2$ and 0.05 M Co(NO$_3$)$_2$, 1 g/l PVP. 
\label{SEM_Figs}}
\end{figure}

A strong dependence of the electrode processes on the solution employed can be easily observed. For the solution containing only Zn ions, the nitrate reduction process starts at approximately -1000~mV vs. SCE (saturated calomel electrode). When other metal ions are present the reduction processes at the working electrode become more complex.  With Co and Cu ions in the solution, we notice that the growth of the nanowires becomes faster and takes place at less electronegative deposition potentials, probably a consequence of the catalytic effect of the metal ions on the nitrate reduction process.

The SEM image in Fig.~\ref{SEM_Figs} displays representative ZnCoO nanowires with cylindrical geometry and 200~nm diameter, as expected. the image reveals that some of the wires are broken during sample preparation. Similar images were obtained for Cu codoped nanowires (not shown here).  By comparing the experimental charge measured during the nanowire growth to the charge estimated from Faraday's law for complete channel filling, we have achieved high pore filling fractions of up to 80$\%$.  The morphology and mechanical stability of the electrodeposited wires depends on the growth rate, which in turn was controlled by the deposition potential and the temperature. In both cases, we observe that nanowires grown at lower deposition rates exhibit a higher mechanical stability, while those grown at higher overpotential broke during the dissolution of the polymer template. 

Two parameters control the composition of the nanowires, namely electrolyte composition and the applied potential. Table~\ref{Compos} summarizes the growth conditions and EDX-determined dopant concentrations of the nanowires. The highest total concentration of (Co,Cu) dopants is found at the lowest negative potential.  The Co content of the nanowires is approximately 4\% for all the nanowires grown, with a weak decrease as the potential becomes more negative. The Cu content has a non-monotonic dependence on potential and varies between 4.6\% and 9.9\%. For potential values more negative than -1000 mV the morphological quality of the wires is low.

\begin{figure} [!]
\includegraphics[width=8cm]{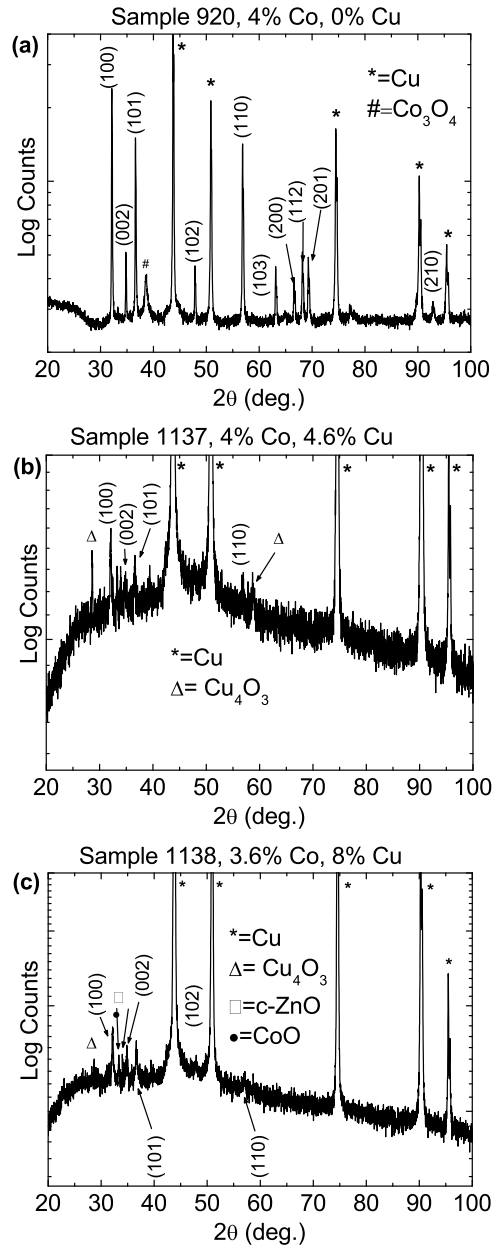}\vspace{+0.4 cm}
\caption{X-ray diffraction spectra for the nanowire arrays of Table~\ref{Compos}. (a) ZnCoO; (b) low Cu codoped wires; (c) moderate Cu codoped wires. Miller indices are for the wurtzite ZnO structure.
\label{XRD_Fig}}
\end{figure}

Precipitate phases form in the nanowires, as are clearly seen in the XRD spectra of Fig.~\ref{XRD_Fig}. Without Cu, the ZnCoO nanowires in Fig.~\ref{XRD_Fig}(a) show a number of sharp peaks corresponding to the wurtzite ZnO structure, as well as a single weak, broad peak from cubic Co$_3$O$_4$.  Figures~\ref{XRD_Fig}(b) and \ref{XRD_Fig}(c) show the XRD spectra of the Cu codoped wires.  In the low Cu codoped nanowires (Fig.~\ref{XRD_Fig}(b)) some of the ZnO peaks have been suppressed, and those that remain are weaker than without Cu codoping.  Additionally, the Co$_3$O$_4$ peak is no longer visible. Instead, (112) and (224) reflections of Cu$_4$O$_3$ have appeared.  At moderate Cu codoping (Fig.~\ref{XRD_Fig}(c)) a CoO peak and a cubic ZnO peak appear and in the high Cu codoped nanowires (not shown) the Cu$_4$O$_3$ and CoO peaks become stronger and sharper.  From these results, the effect of the Cu codoping on the formation of precipitate phases can be summarized as follows: at low Cu concentrations up to 4\%, the Cu reduces the Co$_3$O$_4$ phase, with some of the Cu appearing in the form of Cu$_4$O$_3$. Above 4\% Cu concentration CoO begins to precipitate, and at high Cu concentrations there is a considerable increase in the crystalline order of the secondary Cu$_4$O$_3$ and CoO phases. Using the peak widths of the Co oxide phases, the sizes of the precipitates are calculated from the Scherrer equation as 3.9~nm for Co$_3$O$_4$ in the Co-only nanowires and 2.6~nm for CoO in the moderate Cu codoped nanowires.

\begin{figure}
\includegraphics[width=8cm]{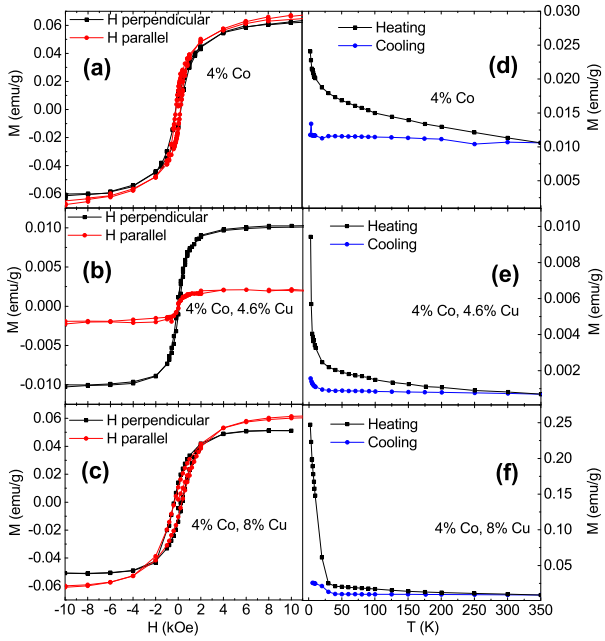}\vspace{+0.0 cm}
\caption{(color online) Magnetization versus field of nanowire arrays measured at 300 K for field applied both perpendicular (black squares) and parallel (red circles) to the nanowire axes for (a) ZnCoO; (b) low Cu codoped ZnCoO; (c) moderate Cu codoped ZnCoO. The component which is linear in H has been subtracted from the plotted data of (a)-(c). Figures (d),(e) and (f) show the remanent magnetization versus temperature measured while heating (black squares) from 2.5 K and subsequent cooling (blue circles) from 350 K for the samples in (a)-(c). 
\label{MH_Co}}
\end{figure}

We now demonstrate the effect of the presence of the Co oxide nano-precipitates on the magnetic properties of the nanowires. Figures~\ref{MH_Co}(a)-(c) show magnetization vs field (M-H) measurements at 300~K.  For Co-only (Fig.~\ref{MH_Co}(a)) and moderate Cu codoped nanowires (Fig.~\ref{MH_Co}(c)), ferromagnetic-like characteristics including saturation, coercivity and remanence are clearly exhibited, with small values of the coercivity (300~Oe), saturation magnetization (M$_S$$\sim$0.07~emu/g) and remanence ratio (M$_R$/M$_S$=0.18), as are typically reported for DMS~\cite{Martin-Gonzalez2008,Kuroda2007}. These results are very similar to the M-H behavior of electrodeposited thin films of ZnCoO~\cite{matei_cobalt-doped_2010} investigated by SQUID magnetometry (not shown) and are also consistent with reports from other groups on Cu-codoped ZnCoO in bulk~\cite{hu2011,liu_enhancement_2013} and thin film~\cite{li_effect_2012,xu_structural_2014} forms.

Although bulk cobalt and copper oxides are antiferromagnetic, nanosized particles of Co$_3$O$_4$ display a weak ferromagnetic character~\cite{Makhlouf2002} that is fully consistent with the magnitude of the ferromagnetic signatures in Fig.~\ref{MH_Co}. As shown in Figs.~\ref{MH_Co}(a) and (c), these nanowires have no anisotropic character to their magnetization, hence the shape anisotropy of the nanowires is not a significant factor. Moreover, the absence of anisotropy is consistent with the origin of the magnetisation being the Co$_x$O$_y$ precipitates detected by XRD in Figs.~\ref{XRD_Fig}(a) and (c). 

By contrast, the M-H loops of low Cu codoped nanowires (Fig.~\ref{MH_Co}(b)) are significantly different, reflecting the lack of Co$_x$O$_y$ phases in these nanowires. M$_S$ is approximately an order of magnitude lower, and although the shape of the loops is ferromagnetic-like, the remanence is barely distinguishable, with a coercive field of less than 100~Oe.  Most dramatically, the low Cu codoped nanowires have a significant magnetic anisotropy.  The easy axis direction is for field perpendicular to the axis of the nanowires, the opposite expected if the nanowire shape was the dominant cause of the anisotropy.  Magnetic anisotropy has been identified before as a key fingerprint of intrinsic (i.e., that does not arise from a parasitic clustered phase) magnetism in high quality crystalline ZnCoO~\cite{Sati2007} and supports the XRD results that indicate these nanowires are largely free of magnetic nano-precipitates.

\begin{figure}
\vspace{-0.6 cm}\includegraphics[width=8cm]{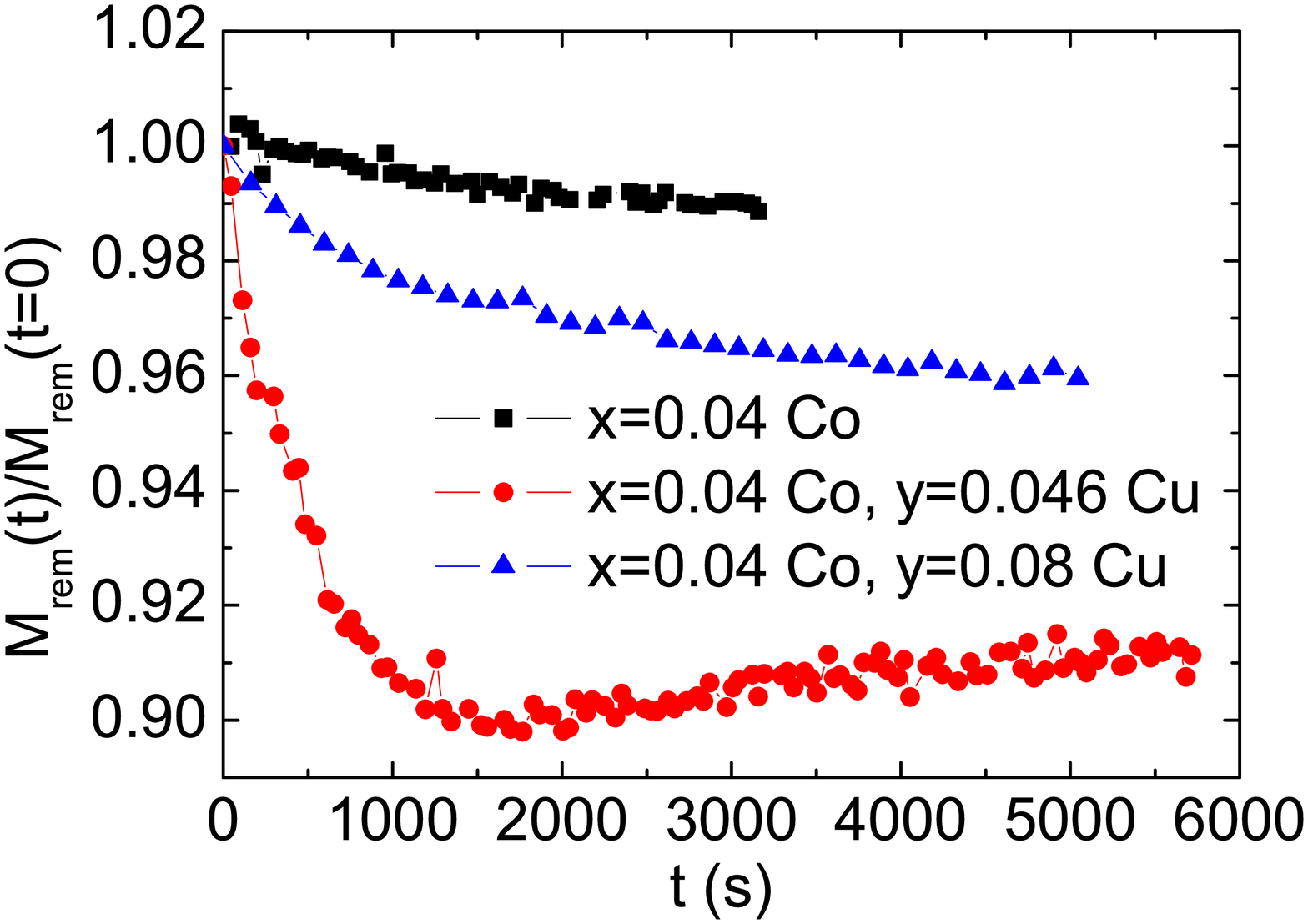} 
\caption{(color online) Thermoremanent magnetization versus time at 2.5 K for nanowires measured after cooling from 300~K in 10~kOe applied field. Data are for Co-only ZnCoO nanowires (black squares), low Cu codoping (red circles) and moderate Cu dodoping (blue triangles).  
\label{Mag_TRM}}
\end{figure}

More evidence for the difference in magnetic properties from the different nano-precipitate phases is seen in Figs.~\ref{MH_Co} (d)-(f), which show the thermoremanent magnetization (TRM) of the nanowires. For these measurements, the nanowires were cooled from 300~K to 2.5~K in a field of 10~kOe, with the field then reduced to zero, and measurements of the magnetization made as a function of time at 2.5~K.  Subsequently, the magnetization was measured whilst heating the sample in zero field to 350~K. A reference Pd sample was used to determine that the true field for the TRM measurements was less than 10 Oe.  As these measurements were carried out in nearly zero field, any interference from paramagnetic transition metal moments, which is often significant in DMS systems, is eliminated, allowing the true cooperative magnetic properties to be investigated.  
 
The nanowires containing Co$_x$O$_y$ precipitates (Figs.~\ref{MH_Co}(d) and (f)) have a TRM $\sim$0.01~emu/g up to at least 350~K whereas the TRM in the Co$_x$O$_y$-free low Cu codoped nanowires (Fig.~\ref{MH_Co}(e)) is an order of magnitude smaller.  The steep drop of TRM in Fig.~\ref{MH_Co}(f) indicates a blocking temperature $T_B$ for the magnetic nano-precipitates of approximately 25-35~K.  Assuming the nano-precipitates are spherical with volume $V$ and using the relation 25$k_BT_B=K_{eff}V$, with the effective magnetic anisotropy constant of Co$_x$O$_y$ nanoparticles $K_{eff}$=6-9$\times$10$^4$~Jm$^{-3}$~\cite{Takada2001,Lagunas2006}, the estimated radius of the nano-precipitates is between 2.8 and 3.6~nm, in agreement with the values derived from the XRD spectra.

A further difference in the magnetization behavior of the nanowires is shown in the time dependent TRM measured at 2.5~K in Fig.~\ref{Mag_TRM}.  The TRM of the Co$_x$O$_y$ nano-precipitate-containing nanowires decays steadily over many thousands of seconds, reminiscent of systems of superparamagnetic nanoparticles with glassy magnetization dynamics~\cite{Sasaki2005}. By contrast, the TRM of the largely Co$_x$O$_y$ precipitate free nanowires has a component that decays over 1000~s and thereafter the magnetization stabilizes. The stability of the TRM after 1000~s indicates that the magnetization of these nanowires is bulk-like, with no more than a 10\% component from nano-precipitates. The fast decay of this weak glassy component indicates any magnetic nano-precipitates in these nanowires have a smaller average size and/or weaker interaction strength than in the nanowires with nano-precipitates detectable by XRD.   

In summary, we have shown that it is possible to control the concentration of Cu codopants in ZnCoO nanowires exclusively by varying the potential used during the electrodeposition process. The formation of Co oxide precipitates in the nanowires can be modified by varying the Cu content between 0 and 8$\%$, thus tuning the magnetic properties. Nanowires containing precipitates have isotropic field loops and a time-dependent magnetization at low temperatures indicating glassy dynamics.  Optimal Cu codoping at near $4\%$ inhibits the formation of the Co oxide nano-precipitates, resulting in nanowires
that exhibit a temporally stable, anisotropic magnetization.  This study validates the conclusion of a theoretical
study~\cite{Zhang2011}, which predicts that Cu dopants in ZnO disaggregate
Co by making the formation of Co-O-Co chains
energetically less favorable. This simple method of
controlling the magnetic characteristics of ZnCoO with
nano-precipitates clears the path for producing material
with optimized properties for high sensitivity magnetic
sensors.

We acknowledge support from the Swiss National Science Foundation (SCOPES project no. IZ73Z0$\textunderscore$127968) and the Romanian Education and Research Ministry UEFISCDI Contracts IDEI 24/2013 and PD 18/2013. 

\end{document}